\documentclass[aps,prx,preprint,superscriptaddress]{revtex4-2}
\usepackage{amsmath}
\usepackage{graphicx}

\begin{document}


\title{On a heuristic point of view concerning the optical activity}


\author{Chun-Fang Li}
\email[Corresponding author: ]{cfli@shu.edu.cn}
\affiliation{Department of Physics, Shanghai University, 99 Shangda Road, 200444 Shanghai, China}

\author{Zhi-Juan Hu}
\email[]{huzhijuan@shnu.edu.cn}
\affiliation{Department of Physics, Shanghai Normal University, 100 Guilin Road, 200233 Shanghai, China}


\date{\today}

\begin{abstract}

Motivated by a recent finding that Fresnel's phenomenological description of the optical activity in the chiral medium is not self-consistent, we conduct a thorough investigation into the nature of the polarization of a plane light wave. We demonstrate that the polarization of light is the reflection of one of its quantum-mechanical properties, called the quasi-spin. Unexpectedly, the quasi-spin is not an observable with respect to the laboratory coordinate system. Instead, it is with respect to the momentum-dependent local coordinate system. The representative operators for the quasi-spin are the Pauli matrices. The wavefunction is the Jones vector. In order to completely determine a state of polarization, two different kinds of degrees of freedom are needed. One is the degrees of freedom to characterize the state of quasi-spin. They are the Stokes parameters, the expectation values of the Pauli matrices in the state described by the Jones vector. The other is the degrees of freedom to specify the local coordinate system, including the propagation direction and an angle of rotation about it. Accordingly, there are two independent mechanisms to change the state of polarization. One is to change the state of quasi-spin in a fixed local coordinate system. This is the traditional mechanism that can be expressed as an SU(2) rotation of the Jones vector. The other is to change the local coordinate system with the state of quasi-spin remaining fixed in it. At last we show that it is the newly-identified mechanism that accounts for the optical activity.

\end{abstract}



\maketitle


\newpage

\section{Introduction}

Optical activity is one of the most fundamental physical phenomena. It refers to the ability of a medium to rotate the orientation of the polarization ellipse of an elliptically polarized light wave that passes through it. Such an ability is associated with the chirality of the molecular or crystalline configuration of the optically-active medium \cite{Barr, Kami}. 
A helically-coiled optical fiber also shows optical activity \cite{Papp-H, Ross, Qian-H, Chen-R}.
It has been widely accepted \cite{Lakh-VV, Bass-PE, Viit-LST, Ghos-F, Pfei-F, Papp-H, Ross, Qian-H, Xi-WWB, Alex-BLY, Chen-R} that optical activity lies with the allogyric birefringence \cite{Ditc} or the circular birefringence \cite{Hecht}; that is, 
the right-handed circularly polarized (RCP) and left-handed circularly polarized (LCP) waves in the chiral medium propagate at different velocities. 
Although the assumption of the existence of circular birefringence, first suggested by Fresnel in 1825, succeeds in describing the rotation of the polarization plane of a linearly polarized wave, it fails to account for the propagation velocity of a linearly polarized wave.
Recently we proved \cite{Hu-L} that circular birefringence does not exist at all in the chiral medium.

The fact is that there exist two orthonormal linearly polarized modes in the chiral medium, just like in the achiral medium. They have the same propagation velocity. 
One can take them as base modes to expand an elliptically polarized wave. As a result, any elliptically polarized wave, including circularly polarized waves, propagates at the velocity of the linearly polarized waves.
The key point is that the two orthonormal linearly polarized waves in the chiral medium are rotated in the same way. Such a rotation is expressed by their rotatory polarization vectors. In fact, it is their rotatory polarization vectors that are orthogonal to each other. When they are taken as base modes to expand an elliptically polarized wave, their polarization vectors serve as the polarization bases of the elliptically polarized wave.
This means that the optical activity amounts to the rotation of the polarization bases, without changing the expansion coefficients that constitute the Jones vector. 
In other words, the optical activity implies a new mechanism to change the polarization of light. It is improper to interpret the optical activity as the change of the Jones vector \cite{Ditc, Dama, Gold}. 
The purpose of the present paper is to reveal the physics behind this mechanism. The contents are arranged as follows.

In order to have a clearer look at the problem that we are faced up with, we briefly review in Section \ref{BR} the main results that we obtained in Ref. \cite{Hu-L} from a slightly different point of view. In particular, we show that what completely describes the state of polarization of a plane wave is its polarization vector rather than its Jones vector. 
In Section \ref{DoF} we prove from the point of view of measurement that two different kinds of degrees of freedom are needed to determine a polarization vector. One is the Stokes parameters that are solely determined by the Jones vector. The other is the degrees of freedom to specify a local coordinate system in which the Stokes parameters as well as the Jones vector are defined, including the propagation direction and an angle of rotation about it.
The Stokes parameters are shown in Section \ref{QS} to be the expectation values of the representative operators
for a quantum-mechanical observable, called the quasi-spin, in a state described by the Jones vector. 
They constitute the so-called Poincar\'{e} vector \cite{Merz} and take the role of characterizing the state of quasi-spin in the local coordinate system.
On this basis, it is argued in Section \ref{PMoP} that the state of polarization is the reflection of the state of quasi-spin in the local coordinate system. 
It not only has the property of the quasi-spin state but also conveys the information of the local coordinate system. Thus there are, in principle, two independent mechanisms to change the state of polarization. One is to change the state of quasi-spin in a fixed local coordinate system. This is the traditional mechanism that can be interpreted in terms of the change of the Jones vector. The other is to change the local coordinate system with the state of quasi-spin remaining fixed in it. 
These two mechanisms are further discussed in Section \ref{understand}. We show that the traditional mechanism is responsible for implementing the change of the polarization state by use of waveplates. But it is the newly-identified mechanism that accounts for the rotation of the polarization bases in the chiral medium.
Section \ref{conclusions} concludes the paper.

\section{Brief review of new description of optical activity}\label{BR}

Chiral objects provide a coupling of the magnetic field to the electric polarizability and of the electric field to the magnetic polarizability. This is described macroscopically by the constitutive relations.
The well-accepted constitutive relations for an isotropic and lossless chiral medium can be written as follows \cite{Silv, Geor},
\begin{subequations}\label{CR}
	\begin{align}
		\mathbf{D} &= \varepsilon \mathbf{E}-g \partial{\mathbf{H}}/\partial{t}, \label{D-H} \\
		\mathbf{B} &= \mu \mathbf{H}+g \partial{\mathbf{E}}/\partial{t},         \label{B-E}
	\end{align}
\end{subequations}
where $\mathbf E$, $\mathbf H$, $\mathbf D$, and $\mathbf B$ are, as usual, the vectors of electric field, magnetic field, electric displacement, and magnetic induction, respectively, $\varepsilon$ is the permittivity, $\mu$ is the permeability, and the pseudo-scalar constant $g$ is the gyrotropic parameter.
Let us first prove that circular birefringence does not exist in the chiral medium.

\subsection{No circular birefringence}

Traditionally, the RCP and LCP plane waves in the chiral medium were referred to as the characteristic modes with different propagation velocities. Suppose that the propagation direction is along the positive $z$-axis. According to Refs. \cite{Ditc, Hecht}, the electric fields of the RCP and LCP plane waves of normalized amplitude and of the same frequency $\omega$ take the form of
\begin{subequations}\label{Er-and-El}
	\begin{align}
		\mathbf{E}_R (z,t) & =\frac{1}{\sqrt 2} (\bar{x}+i\bar{y}) e^{i(k_R z-\omega t)}, \label{Er} \\
		\mathbf{E}_L (z,t) & =\frac{1}{\sqrt 2} (\bar{x}-i\bar{y}) e^{i(k_L z-\omega t)}, \label{El}
	\end{align}
\end{subequations}
where $\bar x$ and $\bar y$ are the unit vectors along the $x$- and $y$-axes, respectively, and $k_R$ and $k_L$ are their ``wave numbers''. 
When constitutive relations (\ref{CR}) are taken into consideration, it can be shown \cite{Silv, Geor} that
\begin{equation}\label{kr-kl}
	\begin{split}
		k_R &=k-\tau,\\
		k_L &=k+\tau,
	\end{split}
\end{equation}
where $k=(\varepsilon \mu)^{1/2} \omega$ and $\tau=-g \omega^2$. It is noted that $\tau$ depends on the gyrotropic parameter $g$; whereas $k$ does not.
Eqs. (\ref{Er}) and (\ref{El}) were commonly interpreted to express two waves that propagate at velocities $\frac{\omega}{k_R}$ and $\frac{\omega}{k_L}$, respectively. The problems with such an interpretation are as follows.

It is true that the coherent superposition of $\mathbf{E}_R$ and $\mathbf{E}_L$ in (\ref{Er-and-El}) can give rise to linearly polarized waves. In particular, the following two superpositions are linearly-polarized,
\begin{equation}\label{E1-E2(0)}
	\begin{split}
		\mathbf{E}_1 &= \frac{\mathbf{E}_R +\mathbf{E}_L}{\sqrt 2} 
		              = (\bar{x} \cos \tau z +\bar{y} \sin \tau z) e^{i(kz-\omega t)}, \\
		\mathbf{E}_2 &= \frac{\mathbf{E}_R -\mathbf{E}_L}{i \sqrt 2} 
		              = (-\bar{x} \sin \tau z +\bar{y} \cos \tau z) e^{i(kz-\omega t)}.
	\end{split}
\end{equation}
Rotated in the same way, they must propagate at the same velocity. To see this more clearly, we rewrite them in the following way,
\begin{equation}\label{E1-E2}
	\begin{split}
		\mathbf{E}_1 &= \bar{x}'(z) \exp[i(kz-\omega t)],\\
		\mathbf{E}_2 &= \bar{y}'(z) \exp[i(kz-\omega t)],
	\end{split}
\end{equation}
where
\begin{equation}\label{PB-linear}
	\begin{split}
		\bar{x}'(z) &=\exp[-i(\bar{z} \cdot \mathbf{\Sigma}) \tau z] \bar{x},\\
		\bar{y}'(z) &=\exp[-i(\bar{z} \cdot \mathbf{\Sigma}) \tau z] \bar{y},
	\end{split}
\end{equation}
$\exp[-i(\bar{z} \cdot \mathbf{\Sigma}) \tau z]$ is the matrix of rotation about the $z$-axis, $\bar z$ is the unit vector along the $z$-axis, and $(\Sigma_k)_{ij} =-i \epsilon_{ijk}$ with $\epsilon_{ijk}$ the Levi-Civit\'{a} pseudotensor.
Nevertheless, if $\mathbf{E}_R$ and $\mathbf{E}_L$ were considered to propagate at different velocities, it would be impossible to deduce from Eqs. (\ref{E1-E2(0)}) that $\mathbf{E}_1$ and $\mathbf{E}_2$ have the same propagation velocity.

As a matter of fact, from Eqs. (\ref{E1-E2(0)}) one readily obtains
\begin{equation}\label{ER-EL(0)}
	\begin{split}
		\mathbf{E}_R &=\frac{1}{\sqrt 2} (\mathbf{E}_1 +i\mathbf{E}_2), \\
		\mathbf{E}_L &=\frac{1}{\sqrt 2} (\mathbf{E}_1 -i\mathbf{E}_2),
	\end{split}
\end{equation}
meaning that whether the RCP or the LCP wave is a coherent superposition of the linearly polarized waves $\mathbf{E}_1$ and $\mathbf{E}_2$. Now that these two linearly polarized waves propagate at the same velocity, the RCP and LCP waves cannot propagate at different velocities. In other words, there is no circular birefringence in the chiral medium. Then what is the propagation velocity of plane waves in the chiral medium?

\subsection{Propagation velocity of plane waves}

When interpreting the RCP wave $\mathbf{E}_R$ in (\ref{Er}) as propagating at velocity $\frac{\omega}{k_R}$, one might imply that it is a superposition of the following two orthogonal linearly polarized waves,
\begin{subequations}\label{LPWs}
	\begin{align}
		\mathbf{E}_1^R &=\bar{x} \exp[i(k_R z-\omega t)], \\
		\mathbf{E}_2^R &=\bar{y} \exp[i(k_R z-\omega t)],
	\end{align}
\end{subequations}
which propagate at velocity $\frac{\omega}{k_R}$. Meanwhile, when interpreting the LCP wave $\mathbf{E}_L$ in (\ref{El}) as propagating at velocity $\frac{\omega}{k_L}$, one might imply that it is a superposition of the following two orthogonal linearly polarized waves,
\begin{equation*}
	\begin{split}
		\mathbf{E}_1^L &=\bar{x} \exp[i(k_L z-\omega t)], \\
		\mathbf{E}_2^L &=\bar{y} \exp[i(k_L z-\omega t)],
	\end{split}
\end{equation*}
which propagate at velocity $\frac{\omega}{k_L}$. But it is impossible for one medium to transmit two sets of linearly polarized waves that, on the one hand, have the same frequency and, on the other hand, propagate at different velocities. The only possibility is that both of these sets of linearly polarized waves do not exist.
Indeed, the polarization of a linearly polarized wave in the chiral medium cannot remain unchanged in transmission. It is instead rotated with propagation as is shown by Eqs. (\ref{E1-E2(0)}) or (\ref{E1-E2}).

As discussed before, the two linearly polarized waves $\mathbf{E}_1$ and $\mathbf{E}_2$ propagate at the same velocity. Since the operator
$\exp[-i(\bar{z} \cdot \mathbf{\Sigma}) \tau z]$
in (\ref{PB-linear}) conveys the rotation of their polarization states, the common phase factor $\exp[i(kz-\omega t)]$ in (\ref{E1-E2}) should be their propagation factor. It determines their phase velocity to be $\frac{\omega}{k}$.
More importantly, these two linearly polarized waves are orthogonal to each other. They form a set of linearly polarized base modes for plane waves propagating along the $z$-axis. 
Their linear combination gives rise to a physically permissible wave,
\begin{equation}\label{EF}
	\mathbf{E} =\alpha_1 \mathbf{E}_1 +\alpha_2 \mathbf{E}_2
	           =\mathbf{a}(z) \exp[i(kz-\omega t)],
\end{equation}
which is in general elliptically polarized, where
\begin{equation}\label{a-vector}
	\mathbf{a}(z) =\alpha_1 \bar{x}'(z) +\alpha_2 \bar{y}'(z)
\end{equation}
and the combination constants satisfy $|\alpha_1|^2 +|\alpha_2|^2 =1$.
In fact, as is shown by Eqs. (\ref{PB-linear}), the unit vectors $\bar{x}'$ and $\bar{y}'$ result from the same rotation of $\bar x$ and $\bar y$, respectively, and are thus orthogonal to each other at the same propagation distance $z$. As physically permissible polarization vectors, they serve as a set of polarization bases \cite{Merz} to expand the polarization vector $\mathbf a$ via Eq. (\ref{a-vector}). The expansion coefficients $\alpha_1$ and $\alpha_2$ constitute the well-known Jones vector \cite{Dama, Gold}
$\alpha =\bigg(\begin{array}{c}
	\alpha_1 \\ \alpha_2
\end{array}
\bigg)$.
Upon substituting Eqs. (\ref{PB-linear}) into Eq. (\ref{a-vector}), we have
\begin{equation}\label{RofPV}
	\mathbf{a}(z) =\exp[-i(\bar{z} \cdot \mathbf{\Sigma}) \tau z] (\alpha_1 \bar{x} +\alpha_2 \bar{y}).
\end{equation}
This indicates that the polarization state of elliptically polarized wave (\ref{EF}) is rotated in the same way as those of linearly polarized waves (\ref{E1-E2}), irrespective of its Jones vector. Consequently, it propagates at the phase velocity $\frac{\omega}{k}$.

\subsection{Polarization vectors of RCP and LCP waves and rotation of polarization bases}

We are now clear that even though the RCP wave in the chiral medium is correctly expressed by Eq. (\ref{Er}), it does not propagate at the velocity $\frac{\omega}{k_R}$. That is to say, $\exp[i(k_R z-\omega t)]$ is not its propagation factor. This is equivalent to say that the RCP wave cannot be resolved into the linearly polarized waves in (\ref{LPWs}) so that the constant unit vectors $\bar x$ and $\bar y$ cannot be regarded as its polarization bases.
All of these means that 
$\frac{1}{\sqrt 2} (\bar{x}+i\bar{y})$ is not the polarization vector of the RCP wave. Similarly,
$\frac{1}{\sqrt 2} (\bar{x}-i\bar{y})$ is not the polarization vector of the LCP wave, either. Let us look at what the polarization vectors of the RCP and LCP waves are.

The polarization vector of any elliptically polarized wave is given by Eq. (\ref{a-vector}). It shows that the rotation of the polarization state comes only from the rotation of the polarization bases $\bar{x}'(z)$ and $\bar{y}'(z)$, without involving the change of the Jones vector $\alpha$.
When
$\alpha= \frac{1}{\sqrt 2} \bigg(\begin{array}{c}
      	                             1 \\ i
                                 \end{array}
                           \bigg)
$,
it becomes the polarization vector of the RCP wave and is given by
\begin{equation}\label{r-vector}
	\mathbf{r}(z) =\frac{1}{\sqrt 2} \exp[-i(\bar{z} \cdot \mathbf{\Sigma}) \tau z]	
	              (\bar{x}+i\bar{y}).
\end{equation}
Likewise, when
$\alpha= \frac{1}{\sqrt 2} \bigg(\begin{array}{c}
      	                             1 \\ -i
                                 \end{array}
                           \bigg)
$,
it becomes the polarization vector of the LCP wave and is given by 
\begin{equation}\label{l-vector}
    \mathbf{l}(z) =\frac{1}{\sqrt 2} \exp[-i(\bar{z} \cdot \mathbf{\Sigma}) \tau z]
                  (\bar{x}-i\bar{y}).
\end{equation}
Indeed, because $\bar{x}+i \bar{y}$ and $\bar{x}-i \bar{y}$ are eigen vectors of matrix $\bar{z} \cdot \mathbf{\Sigma}$ with eigenvalues $+1$ and $-1$, respectively, the above two equations can be re-expressed in the following way,
\begin{equation}\label{r-l}
	\begin{split}
		\mathbf{r}(z) &=\frac{1}{\sqrt 2} (\bar{x}+i\bar{y}) \exp(-i\tau z), \\
		\mathbf{l}(z) &=\frac{1}{\sqrt 2} (\bar{x}-i\bar{y}) \exp(i \tau z). 
	\end{split}
\end{equation}
With the help of them, we can rewrite expressions (\ref{Er-and-El}) as
\begin{equation}\label{Er'-El'}
	\begin{split}
		\mathbf{E}_R (z,t) & =\mathbf{r}(z) \exp[i(kz-\omega t)],\\
		\mathbf{E}_L (z,t) & =\mathbf{l}(z) \exp[i(kz-\omega t)],
	\end{split}
\end{equation}
by virtue of Eqs. (\ref{kr-kl}).
From these discussions we conclude that the polarization state of a circularly polarized wave, as a special case of elliptically polarized wave (\ref{EF}), is not unchanged in transmission. It is rotated with propagation in the same way as those of linearly polarized waves (\ref{E1-E2}). The only difference is that the result of rotation in such a case appears as a $z$-dependent phase factor. 
That is to say, the $z$-dependent phase factors in (\ref{r-l}) convey a dependence of the polarization state on the propagation distance. 
The phases acquired by the RCP and LCP waves over the same propagation distance are the same in magnitude but are opposite in sign. This may help to understand why the RCP and LCP waves were incorrectly interpreted to propagate at different phase velocities in traditional description.

It is important to point out that circular polarization vectors (\ref{r-vector}) and (\ref{l-vector}) are also orthogonal to each other at the same propagation distance. They can serve as a set of polarization bases, too. In terms of them, polarization vector (\ref{a-vector}) can be expanded as follows,
\begin{equation}\label{a-CPB}
	\mathbf{a}=\frac{1}{\sqrt 2}(\alpha_1 -i\alpha_2) \mathbf{r}
	+\frac{1}{\sqrt 2}(\alpha_1 +i\alpha_2) \mathbf{l}.
\end{equation}
It shows that the rotation of the circular polarization bases $\mathbf r$ and $\mathbf l$ is responsible for the rotation of the polarization state of the elliptically polarized wave when considered as a coherent superposition of the RCP and LCP waves. 
In a word, even though the state of polarization described by polarization vector (\ref{a-vector}) varies with the propagation distance $z$, the Jones vector $\alpha$ does not. The variation of the state of polarization amounts to the variation of the polarization bases $\bar{x}'(z)$ and $\bar{y}'(z)$ with the propagation distance.

\section{Degrees of freedom for the state of polarization}\label{DoF}

Traditionally, the Jones vector was believed \cite{Gold} to take the role of describing the polarization of light. But the above discussions about the optical activity suggest that this may not be the case. 
What completely describes the polarization state of a plane wave can only be its polarization vector. More importantly, as is shown by Eqs. (\ref{r-l}), the change of polarization state under some particular conditions appears as a phase factor of the polarization vector. That is to say, polarization vectors that differ by phase factors may not represent the same polarization state.
This naturally raises the question about the \emph{nature} of the polarization of light. It is generally thought that the experimentally accessible quantities to characterize the state of polarization are the Stokes parameters \cite{Hecht, Fano, Hecht-70, Gold, Jauc-R}. The problem, however, is that what directly determines the Stokes parameters is the Jones vector \cite{Dama} rather than the polarization vector. Now that the Jones vector is not able to completely describe the polarization, the Stokes parameters are not able to completely characterize the polarization, either. 
To truly understand the nature of the polarization, it is crucial to figure out the degrees of freedom that determine the state of polarization. Let us first make use of the polarization vector of a plane wave in a dielectric medium, which is achiral, to analyze how many degrees of freedom are needed to determine a state of polarization.

\subsection{Number of degrees of freedom}

For the sake of clarity, the plane wave is assumed to propagate in an arbitrary direction denoted by $\mathbf w$, the unit wavevector. The electric field of it can be written in the form: 
\begin{equation}\label{E-a}
    \mathbf{E}(\mathbf{x},t)=\mathbf{a} E \exp[i(k\mathbf{w} \cdot \mathbf{x}-\omega t)],
\end{equation}
where $\mathbf a$ is the polarization vector obeying the normalization condition
\begin{equation}\label{normalization}
    \mathbf{a}^* \cdot \mathbf{a}=1,
\end{equation}
$E$ is the amplitude, and $k$ is the wave number. What should be kept in mind is that the polarization vector has nothing to do with the phase coming from the propagation factor
$\exp[i(k\mathbf{w} \cdot \mathbf{x}-\omega t)]$.
It has three Cartesian components,
\[ \mathbf{a}=      a_x \bar{x}+ a_y \bar{y}+ a_z \bar{z}
             \equiv \Bigg(\begin{array}{c}
                            a_x \\
                            a_y \\
                            a_z
                          \end{array}
                    \Bigg). \]
Each of them is a complex-valued number. 
So after normalization condition (\ref{normalization}) is taken into consideration, there are five independent real-valued parameters or, equivalently, five degrees of freedom to determine the polarization vector. 
However, the polarization vector has to be perpendicular to the propagation direction $\mathbf w$,
\begin{equation*}
    \mathbf{w} \cdot \mathbf{a}=0,
\end{equation*}
in accordance with the Maxwell's equation $\nabla \cdot \mathbf{E}=0$. 
This means that the two independent real-valued parameters specifying the propagation direction $\mathbf w$ are two degrees of freedom to determine a state of polarization. As a result, apart from $\mathbf w$, three degrees of freedom are still needed. 

\subsection{Stokes parameters contain only two degrees of freedom}

Now we pay attention to the Stokes parameters. As mentioned above, it is impossible to determine the Stokes parameters directly from the polarization vector. To do this, one has to introduce the Jones vector.
Let be $\mathbf u$ and $\mathbf v$ two mutually-perpendicular unit vectors that, with $\mathbf w$, form a right-handed Cartesian coordinate system,
\begin{equation}\label{triad}
    \mathbf{u} \cdot \mathbf{u}=\mathbf{v} \cdot \mathbf{v}=1, \quad 
    \mathbf{u} \cdot \mathbf{v}=0, \quad \mathbf{u} \times \mathbf{v} =\mathbf{w}.
\end{equation}
Being perpendicular to $\mathbf w$, they represent two orthogonal polarization states that are physically permissible in the achiral medium.
When they are taken as the polarization bases to expand the polarization vector,
\begin{equation}\label{a-expand}
    \mathbf{a}=\alpha_1 \mathbf{u} +\alpha_2 \mathbf{v},
\end{equation}
the expansion coefficients constitute the Jones vector 
$\alpha =\bigg(\begin{array}{c}
           	       \alpha_1 \\ \alpha_2
               \end{array}
         \bigg)$,
which is normalized
\begin{equation}\label{unit-alpha}
    \alpha^\dag \alpha =1
\end{equation}
in accordance with Eq. (\ref{normalization}). Upon expressing the polarization bases as three-component column matrices
$\mathbf{u}=\Bigg(\begin{array}{c}
                      u_x \\ u_y \\ u_z
                  \end{array}
            \Bigg)
$
and
$\mathbf{v}=\Bigg(\begin{array}{c}
                      v_x \\ v_y \\ v_z
                  \end{array}
            \Bigg),
$
we can convert expansion (\ref{a-expand}) into
\begin{equation}\label{a-alpha}
    \mathbf{a}=\varpi \alpha,
\end{equation}
where
$\varpi=(\begin{array}{lr}
             \mathbf{u} & \mathbf{v}
         \end{array})
$
is a $3 \times 2$ matrix. The matrix $\varpi$ satisfies
\begin{equation}\label{unitarity1}
	\varpi^\dag \varpi =I_2
\end{equation}
by virtue of Eqs. (\ref{triad}), where $I_2$ is the unit $2 \times 2$ matrix. Multiplying (\ref{a-alpha}) by $\varpi^\dag$ on the left and making use of (\ref{unitarity1}), we get
\begin{equation}\label{alpha-a}
	\alpha =\varpi^\dag \mathbf{a}.
\end{equation}
In terms of the Jones vector, the Stokes parameters are expressed as
\begin{equation}\label{SP}
    s_i =\alpha^\dag \hat{\sigma}_i \alpha, \quad i=1,2,3,
\end{equation}
where
\begin{equation}\label{PM}
	\hat{\sigma}_1=\bigg(\begin{array}{cc}
	     	          1 &  0 \\
		              0 & -1
	                     \end{array}
	               \bigg),                \quad
	\hat{\sigma}_2=\bigg(\begin{array}{cc}
		              0 & 1 \\
		              1 & 0
	                     \end{array}
	               \bigg),                \quad
	\hat{\sigma}_3=\bigg(\begin{array}{cc}
		              0 & -i \\
		              i &  0
	                     \end{array}
	               \bigg)
\end{equation}
are the Pauli matrices. 

Of course, if one chooses circular polarization bases \cite{Merz}, which are given by
\begin{eqnarray*}
	\mathbf{R} &=& \frac{1}{\sqrt 2} (\mathbf{u} +i\mathbf{v}), \\
	\mathbf{L} &=& \frac{1}{\sqrt 2} (\mathbf{u} -i\mathbf{v}),
\end{eqnarray*}
the polarization vector can be expanded as follows,
\[ \mathbf{a} = \alpha_R \mathbf{R} + \alpha_L \mathbf{L} \equiv \varpi^c \alpha^c ,\]
where
$
\alpha_R=\frac{1}{\sqrt 2}(\alpha_1 -i\alpha_2)
$,
$
\alpha_L=\frac{1}{\sqrt 2}(\alpha_1 +i\alpha_2)
$,
$
\varpi^c =(\begin{array}{lr}
               \mathbf{R} & \mathbf{L}
           \end{array})
$, and 
$
\alpha^c =\bigg(\begin{array}{c}
                	\alpha_R \\ \alpha_L
                \end{array}
          \bigg)  
$.
In this case, the Stokes parameters (\ref{SP}) can be rewritten as
\[ s_i = \alpha^{c\dag} \hat{\sigma}^c_i \alpha^c ,\]
where
\begin{equation*}
	\hat{\sigma}^c_1=\bigg(\begin{array}{cc}
		                      0 & 1 \\
		                      1 & 0
	                       \end{array}
	                 \bigg),                \quad
	\hat{\sigma}^c_2=\bigg(\begin{array}{cc}
		                      0 & -i \\
		                      i &  0
	                       \end{array}
	                 \bigg),                \quad
	\hat{\sigma}^c_3=\bigg(\begin{array}{cc}
	                          1 &  0 \\
	                          0 & -1
                           \end{array}
                     \bigg).            
\end{equation*}
The circular polarization bases $\varpi^c$ are related to the linear polarization bases $\varpi$ through
\begin{equation}\label{ToPB}
	\varpi^c =\varpi M^\dag ,
\end{equation} 
where
$
M=\frac{1}{\sqrt 2} \bigg(\begin{array}{cc}
                        	1 & -i \\
                        	1 &  i
                          \end{array}
                    \bigg),
$
so that
$\alpha^c =M \alpha$ and
$\hat{\sigma}^c_i =M \hat{\sigma}_i M^\dag$. The transition from the linear polarization bases to the circular polarization bases does not affect the Stokes parameters. 
In this paper, we will only use the linear polarization bases.

According to normalization condition (\ref{unit-alpha}), the Stokes parameters satisfy
\begin{equation*}
	s_1^2 +s_2^2 +s_3^2 =1,
\end{equation*}
meaning that only two of the three Stokes parameters are independent. In other words, only two degrees of freedom are contained in the Stokes parameters. We are thus convinced that the Stokes parameters, together with the propagation direction $\mathbf w$, are not enough to completely determine a state of polarization. One additional degree of freedom is needed.

\subsection{The fifth degree of freedom and local coordinate system}\label{NDoF}

To find out the additional degree of freedom, we reexamine the Jones vector that fully determines the Stokes parameters.
It is noticed that the transverse axes that are taken to introduce the Jones vector via (\ref{alpha-a}) are indeterminate to the extent that a rotation about the propagation direction can be performed \cite{Mand-W}. By this it is meant that the Jones vector is an entity in association with the transverse axes. For a given polarization vector, the Jones vector associated with different transverse axes should be different. 
Denote by
$\varpi' =(\begin{array}{cc}
	         \mathbf{u}' & \mathbf{v}'
           \end{array})
$
the new transverse axes that result from a rotation of the old ones, $\varpi$, by an angle $\phi$,
\begin{equation}\label{rotation-bases}
    \varpi' =\exp[-i(\mathbf{w} \cdot \mathbf{\Sigma}) \phi] \varpi.
\end{equation}
Taking the primed unit vectors $\mathbf{u}'$ and $\mathbf{v}'$ as new polarization bases, the same polarization vector that is expanded in terms of the old polarization bases $\varpi$ via (\ref{a-alpha}) can be expanded as
\begin{equation*}
	\mathbf{a}=\alpha'_1 \mathbf{u}' +\alpha'_2 \mathbf{v}' =\varpi' \alpha',
\end{equation*}
where the expansion coefficients $\alpha'_1$ and $\alpha'_2$ constitute the new Jones vector
$\alpha' =\bigg(\begin{array}{c}
	                \alpha'_1 \\ \alpha'_2
                \end{array}
          \bigg)$.
Since the new polarization bases also satisfies
$ \varpi'^\dag \varpi' =I_{2} $,
we multiply the above equation by $\varpi'^\dag$ on the left and take this relation into account to give
\begin{equation}\label{alpha'-a}
	\alpha' =\varpi'^\dag \mathbf{a}.
\end{equation}
Upon substituting Eqs. (\ref{rotation-bases}) and (\ref{a-alpha}), we finally get 
\begin{equation}\label{trans-JV}
	\alpha' =\exp(i \hat{\sigma}_3 \phi) \alpha,
\end{equation}
where the equality
\[ \hat{\sigma}_3 =\varpi^\dag (\mathbf{w} \cdot \mathbf{\Sigma}) \varpi \]
has been used. Eq. (\ref{trans-JV}) constitutes the transformation of the Jones vector of the given polarization vector $\mathbf a$ under the transformation (\ref{rotation-bases}) of the transverse axes. It indicates that the new Jones vector cannot be the same as the old one unless $\phi=0$ or $\phi=2\pi$. 
This suggests that the Jones vector as a two-component entity is not defined in the same coordinate system as the polarization vector. 
If the polarization vector is said to be defined in the laboratory coordinate system, the Jones vector cannot be. 

To discern the coordinate system in which the Jones vector is defined, we turn our attention back to the Stokes parameters. It is usually thought \cite{Dama, Merz} that the Stokes parameters are described in an abstract three-dimensional space, constituting a vector in that space. 
However, since the new polarization bases $\varpi'$ are on the same footing as the old ones $\varpi$, it follows from Eq. (\ref{SP}) that the Stokes parameters determined by the new Jones vector $\alpha'$ are given by
\begin{equation}\label{SP'}
	s'_i =\alpha'^\dag \hat{\sigma}_i \alpha', \quad i=1,2,3.
\end{equation}
Now that $\alpha'$ is different from $\alpha$, the new Stokes parameters in (\ref{SP'}) are in general not the same as the old ones in (\ref{SP}). 
In fact, substituting (\ref{trans-JV}) into (\ref{SP'}) and making use of (\ref{SP}), we get
\begin{equation}\label{trans-SP}
	\begin{split}
		s'_1 &= s_1 \cos 2\phi +s_2 \sin 2\phi, \\
		s'_2 &=-s_1 \sin 2\phi +s_2 \cos 2\phi, \\
		s'_3 &=s_3.
	\end{split}
\end{equation}
Because it is impossible for a given state of polarization to have different sets of Stokes parameters in one and the same space, the three-dimensional space in which the Stokes parameters are described cannot be an abstract space. The Stokes parameters must be defined in some concrete coordinate system. Different sets of Stokes parameters are defined in different coordinate systems.
Considering that they are solely determined by the Jones vector, the Stokes parameters should be defined in the same coordinate system as the Jones vector. 
What is more important is that transformation (\ref{trans-JV}) of the Jones vector has a one-to-one correspondence with transformation (\ref{rotation-bases}) of the transverse axes. In particular, when $\phi=\pi$ (or $\varpi' =-\varpi$), one has $\alpha' =-\alpha$. And when $\phi=2\pi$ (or $\varpi' =\varpi$), one has $\alpha' =\alpha$. 
Thus, we have reason to believe that the Jones vector as well as the Stokes parameters is defined in the coordinate system formed by the associated transverse axes and the propagation direction $\mathbf w$. 
This is, of course, not the laboratory coordinate system in which the polarization vector is defined. It is a local coordinate system dependent on the propagation direction.
Apparently, if the local coordinate system is not specified, it will be physically meaningless to talk about a Jones vector or a set of Stokes parameters.
From this one may infer that the degree of freedom that we are searching for should have something to do with the local coordinate system in which the Jones vector and the Stokes parameters are defined.

Mathematically, the Jones vector is related to the polarization vector by Eq. (\ref{alpha-a}). If the Jones vector is defined in the local coordinate system that depends on the propagation direction, then Eq. (\ref{alpha-a}) means that the mapping matrix $\varpi^\dag$ is enough to represent the local coordinate system in which it is defined. 
This is indeed the case. As is indicated by Eqs. (\ref{triad}), the transverse axes $\mathbf u$ and $\mathbf v$ must be connected with some particular propagation direction $\mathbf w$. So the matrix $\varpi^\dag$ that consists of $\mathbf u$ and $\mathbf v$ already conveys the information of $\mathbf w$ in an appropriate way. 
But on the other hand, Eq. (\ref{rotation-bases}) shows that given $\mathbf w$, different local coordinate systems are related to one another by a rotation about it. That is to say, the rotation angle $\phi$ in (\ref{rotation-bases}) is the degree of freedom to distinguish different local coordinate systems that all depend on the same $\mathbf w$.
Now that the Stokes parameters are quantities in the local coordinate system, for the Stokes parameters to be physically meaningful, the local coordinate system has to be specified unambiguously. The rotation angle $\phi$ is therefore the fifth degree of freedom to determine the state of polarization of a plane wave. It combines with the propagation direction $\mathbf w$ to specify the local coordinate system.
It is noted, by the way, that the rotation of the local coordinate system, Eq. (\ref{rotation-bases}), is not to be confused with the transition of the polarization bases, Eq. (\ref{ToPB}), which does not involve the change of the local coordinate system.

In concluding this section we summarize our main results.
In order to completely determine the state of polarization of a plane wave, two different kinds of degrees of freedom are needed. One is the Stokes parameters that contain two degrees of freedom. The other is the degrees of freedom to specify the local coordinate system in which the Stokes parameters are defined, including the propagation direction and an angle of rotation about it. The reason for the Stokes parameters not to be able to completely characterize the polarization is that they are quantities in the local coordinate system whereas the polarization is a phenomenon in the laboratory coordinate system. Let us consider what on earth these experimentally accessible quantities really mean.

\section{Quasi-spin in local coordinate system}\label{QS}

It is well known in quantum mechanics that the Pauli matrices are representative operators for the electron spin. The wavefunction to describe the state of electron spin is a two-component spinor. 
So an analogy with the spin of the electron suggests that the Stokes parameters in (\ref{SP}) correspond to an observable the representative operators for which are the Pauli matrices (\ref{PM}). They are expectation values of these operators in a state described by the wavefunction, the Jones vector $\alpha$.
An important property of the Pauli matrices is that they satisfy the commutation relation of the angular momentum,
\begin{equation}\label{su2}
	[\hat{\sigma}_i, \hat{\sigma}_j]=2i \sum_k \epsilon_{ijk} \hat{\sigma}_k, \quad
	i,j,k=1,2,3,
\end{equation}
except for a factor two. 
In addition, as discussed above, the Stokes parameters constitute a three-dimensional unit vector in the associated local coordinate system. In fact, transformations (\ref{trans-SP}) indicate that $s_3$ is the component of the vector along the propagation direction $\mathbf w$ and, hence, $s_1$ and $s_2$ are the components along the transverse axes $\mathbf u$ and $\mathbf v$, respectively,
\begin{equation}\label{PV}
	\mathbf{P} = s_1 \mathbf{u} +s_2 \mathbf{v} +s_3 \mathbf{w}.
\end{equation}
According to Merzbacher \cite{Merz}, we will call $\mathbf P$ the Poincar\'{e} vector. Substituting Eq. (\ref{SP}) into Eq. (\ref{PV}), we have
\begin{equation*}
	\mathbf{P}=\alpha^\dag \hat{\boldsymbol \sigma} \alpha,
\end{equation*}
where
\begin{equation}\label{QSO}
	\hat{\boldsymbol \sigma}
	=\hat{\sigma}_1 \mathbf{u} +\hat{\sigma}_2 \mathbf{v} +\hat{\sigma}_3 \mathbf{w}.
\end{equation}
Eq. (\ref{QSO}) shows that similar to the spin of the electron, the observable represented by the Pauli matrices here is also a vector. But unexpectedly, its quantization axis is with respect to the local coordinate system. That is to say, it is quantized with respect to the local coordinate system.
In this sense we say that it is the quantum-mechanical property of light in the local coordinate system. For these reasons, we will refer to this observable as the quasi-spin of the photon.

It is not difficult to prove \cite{Dama} that the wavefunction of the quasi-spin, the Jones vector $\alpha$, is an eigenfunction of the unitary unimodular matrix
$\mathbf{P} \cdot \hat{\boldsymbol \sigma}$:
\begin{equation}\label{EVE}
	\mathbf{P} \cdot \hat{\boldsymbol \sigma} \alpha =\alpha.
\end{equation}
Hence, the Poincar\'{e} vector $\mathbf P$ may legitimately be said to point in the direction of the quasi-spin in the local coordinate system. In other words, the Poincar\'{e} vector, or equivalently, the Stokes parameters, takes the role to characterize the state of quasi-spin in the local coordinate system. 
This, in turn, shows that Jones vectors that differ by phase factors describe the same state of quasi-spin, the same as is demonstrated by the bilinear dependence of the Stokes parameters (\ref{SP}) on the Jones vector. 

Algebraic relation (\ref{su2}) means that the operator $\hat{\boldsymbol \sigma}$ for the quasi-spin is the generator of rotation in the \emph{local coordinate system}. This is to say that the operator
$\exp (-\frac{i}{2} \mathbf{n} \cdot \boldsymbol{\sigma} \chi)$
induces a rotation about the direction $\mathbf n$ by an angle $\chi$, where
$\mathbf{n} =n_1 \mathbf{u} +n_2 \mathbf{v} +n_3 \mathbf{w}$ 
is a unit vector in the local coordinate system. It changes the quasi-spin state $\alpha$ into
\begin{equation}\label{rotation-JV}
	\alpha'' =\exp \Big( -\frac{i}{2} \mathbf{n} \cdot \boldsymbol{\sigma} \chi \Big) \alpha.
\end{equation}
So the Jones vector behaves in the local coordinate system as a spinor \cite{Jauc-R, Merz, Berr-JMO,Comm1}. After the rotation, the Poincar\'{e} vector of the quasi-spin state $\alpha''$ is given by
\begin{equation*}
	\mathbf{P}'' = \alpha''^\dag \hat{\boldsymbol \sigma} \alpha''
	             = s''_1 \mathbf{u} +s''_2 \mathbf{v} +s''_3 \mathbf{w},
\end{equation*}
where
$s''_i =\alpha''^\dag \hat{\sigma}_i \alpha''$.
It is easy \cite{Dama,Saku} to prove that if letting
$\mathbf{P} =\Bigg(\begin{array}{c}
	                  s_1 \\ s_2 \\ s_3
                   \end{array}
             \Bigg)$
and
$\mathbf{P}'' =\Bigg(\begin{array}{c}
                        s''_1 \\ s''_2 \\ s''_3
                     \end{array}
               \Bigg)$,
one will have
\begin{equation}\label{RoPV}
	\mathbf{P}'' = \exp[-i(\mathbf{n} \cdot \mathbf{\Sigma}) \chi] \mathbf{P},
\end{equation}
showing that the Stokes parameters indeed transform as the Cartesian components of a vector under rotations in the local coordinate system.

In a word, as revealed by the analyses of the phenomenon of optical activity that we conducted in Section \ref{BR}, the traditional theory for describing the polarization of light is incomplete. 
What the Jones vector truly describes is not the polarization. Instead, it describes the quantum-mechanical property of light in the local coordinate system, the quasi-spin. Such a property has profound implications for the nature of the polarization as we will show below.

\section{Physical meaning of polarization}\label{PMoP}

The polarization vector for the state of polarization is related to the Jones vector for the state of quasi-spin by Eq. (\ref{a-alpha}). 
What is interesting is, as we have seen before, that the matrix $\varpi$ in Eq. (\ref{a-alpha}) elegantly represents the local coordinate system with respect to which the quasi-spin is quantized. 
Let us make use of the plane wave (\ref{E-a}) in the achiral medium to examine in detail how the state of polarization depends on the state of quasi-spin in the local coordinate system.

In the first place, we examine the dependence of the state of polarization on the state of quasi-spin in a fixed local coordinate system. Assume that the local coordinate system $\varpi$ involved in Eqs. (\ref{a-alpha})-(\ref{alpha-a}) is fixed.
Substituting Eq. (\ref{alpha-a}) into Eq. (\ref{a-alpha}) and considering the arbitrariness of $\mathbf a$, we get
\begin{equation}\label{unitarity2}
	\varpi \varpi^\dag =I_3 ,
\end{equation}
where $I_3$ is the unit $3 \times 3$ matrix. Mathematically, Eq. (\ref{unitarity2}) in combination with Eq. (\ref{unitarity1}) means that the matrix $\varpi$ is quasi-unitary \cite{Golu-L}. $\varpi^\dag$ is the Moore-Penrose pseudo inverse of $\varpi$, and vice versa. 
The quasi-unitarity of $\varpi$ guarantees a one-to-one correspondence between the Jones vector $\alpha$ and the polarization vector $\mathbf a$ via Eq. (\ref{a-alpha}) or (\ref{alpha-a}). Hence, to each quasi-spin state $\alpha$ in the fixed local coordinate system there corresponds a unique polarization state $\mathbf a$ in the laboratory coordinate system. 
In this case, the Stokes parameters (\ref{SP}) that characterize the state of quasi-spin can be used to differentiate between different states of polarization.
Such a correspondence provides a mechanism to change the state of polarization, which is to vary the state of quasi-spin in a fixed local coordinate system. This is the traditional mechanism that can be interpreted in terms of the variation of the Jones vector \cite{Dama}.
It should be pointed out that if the $z$-axis of the laboratory coordinate system is taken as the propagation direction, the $x$- and $y$-axes can also be chosen as the transverse axes of the local coordinate system. 
This is the local coordinate system that is commonly chosen in the literature \cite{Gold} to define the Jones vector.
The measurement of the corresponding Stokes parameters therefore should be understood to be done in so chosen local coordinate system.
But such a choice makes it difficult to distinguish the local coordinate system for the quasi-spin from the laboratory coordinate system for the polarization.

In the second place, we examine the dependence of the state of polarization on the local coordinate system in which the state of quasi-spin is fixed. 
To this end, we consider a polarization state the Jones vector of which is the same as that of polarization state (\ref{a-alpha}) but is defined in the primed local coordinate system $\varpi'$,
\begin{equation}\label{a'-alpha}
	\mathbf{a}' =\varpi' \alpha.
\end{equation}
According to Eq. (\ref{rotation-bases}), polarization state (\ref{a'-alpha}) is different from polarization state (\ref{a-alpha}) unless $\phi=0$ or $\phi=2\pi$. 
The rotation angle $\phi$ for the local coordinate system is now the degree of freedom to differentiate between different polarization states. The Stokes parameters can no longer be able to play the role.

We stress that the Stokes parameters constitute a vector, the Poincar\'{e} vector (\ref{PV}), in the local coordinate system. 
But it should be noted that $\mathbf u$, $\mathbf v$, and $\mathbf w$ for the Cartesian axes of the local coordinate system are all unit vectors in the laboratory coordinate system. In this regard, the Poincar\'{e} vector can also be ``viewed'' as a vector in the laboratory coordinate system. Even so, it is not able to completely characterize the state of polarization, the same as the Stokes parameters. 
We have seen in Part \ref{NDoF} that a given state of polarization corresponds to different states of quasi-spin in different local coordinate systems \cite{Comm2} as is shown by Eqs. (\ref{alpha-a}) and (\ref{alpha'-a}). 
The Poincar\'{e} vector in the unprimed local coordinate system $\varpi$ is given by expression (\ref{PV}). Accordingly, the Poincar\'{e} vector in the primed local coordinate system $\varpi'$ is given by
\[ \mathbf{P}' =s'_1 \mathbf{u}' +s'_2 \mathbf{v}' +s'_3 \mathbf{w}, \]
where $s'_i$ ($i=1,2,3$) are defined by (\ref{SP'}). Upon substituting Eqs. (\ref{trans-SP}) and taking Eq. (\ref{rotation-bases}) into consideration, it is not difficult to find
\[ \mathbf{P}' = \exp[i(\mathbf{w} \cdot \mathbf{\Sigma}) \phi] \mathbf{P}, \]
showing that the Poincar\'{e} vectors in different local coordinate systems, when viewed in the laboratory coordinate system, are in general not the same. This proves our assertion.
Ultimately, as is shown by Eq. (\ref{EVE}), what the Poincar\'{e} vector characterizes is the state of quasi-spin in the local coordinate system.

To conclude, the polarization of light described by the polarization vector is the reflection of the quasi-spin of the photon in its local coordinate system. It not only has the property of the quasi-spin but also conveys the information of the local coordinate system in which the quasi-spin is characterized.
The reason for the traditional theory for describing the polarization of light to be incomplete is that it simply
identifies the polarization in the laboratory coordinate system with the quasi-spin in some specific local coordinate system \cite{Saku}.
This is why it could not correctly describe the phenomenon of optical activity in the chiral medium.
The dependence of the polarization on the local coordinate system provides a new mechanism to change the polarization. This is to vary the local coordinate system with the state of quasi-spin remaining fixed in it.
We will see in the next section that it is this newly identified mechanism that underlies the variation of the polarization bases in the chiral medium.

\section{A new mechanism to change the state of polarization}\label{understand}

To better understand the new mechanism for changing the polarization, it is helpful to first interpret the traditional mechanism in terms of the quasi-spin. 
As discussed above, the state of quasi-spin in a fixed local coordinate system has a one-to-one correspondence with the state of polarization in the laboratory coordinate system. So varying the state of quasi-spin in a fixed local coordinate system is able to change the state of polarization. Such a variation can be mathematically expressed as a rotation of the Jones vector as is shown by (\ref{rotation-JV}). 
Suppose that the marix $\varpi^\dag$ in Eq. (\ref{alpha-a}) represents the fixed local coordinate system. If the quasi-spin state $\alpha''$ is different from the quasi-spin state $\alpha$, the polarization state
\begin{equation*}
	\mathbf{a}'' =\varpi \alpha''
\end{equation*} 
will be different from polarization state (\ref{a-alpha}). 
This mechanism can be experimentally implemented by applying waveplates. The fast and slow axes of a waveplate take the role of choosing the transverse axes of the local coordinate system in which the Jones vector is varied \cite{Gold}. In that case, Eq. (\ref{alpha-a}) can be interpreted as mapping the polarization vector $\mathbf a$ onto the Jones vector $\alpha$ in the chosen local coordinate system that is represented by $\varpi^\dag$. 
Upon substituting Eq. (\ref{rotation-JV}) into the above equation and considering Eq. (\ref{alpha-a}), one has
\begin{equation}\label{a''}
	\mathbf{a}'' =\varpi \exp \Big( -\frac{i}{2} \mathbf{n} \cdot \hat{\boldsymbol \sigma} \chi \Big) 
	\varpi^\dag \mathbf{a}.
\end{equation}

Now that the quasi-spin is a quantum-mechanical property with respect to the local coordinate system, varying the local coordinate system with the state of quasi-spin remaining fixed in it is also able to change the state of polarization. 
For instance, if rotating the local coordinate system about the propagation direction without varying the state of quasi-spin in it, one will change the state of polarization.
This is demonstrated by the difference between Eqs. (\ref{a'-alpha}) and (\ref{a-alpha}). In fact, substituting Eq. (\ref{rotation-bases}) into Eq. (\ref{a'-alpha}) and taking Eq. (\ref{a-alpha}) into consideration, one gets
\begin{equation}\label{a'-a}
	\mathbf{a}' =\exp[-i(\mathbf{w} \cdot \mathbf{\Sigma}) \phi] \mathbf{a},
\end{equation} 
which shows that the rotation angle $\phi$ here is a continuous variable to determine the state of polarization.
What we encountered in Section \ref{BR} is precisely such a physical process. Indeed, from Eq. (\ref{RofPV}) we have 
\[ \mathbf{a}(0)=\alpha_1 \bar{x} +\alpha_2 \bar{y}. \]
As a result, we can rewrite Eq. (\ref{RofPV}) as follows,
\[ \mathbf{a}(z) =\exp[-i(\bar{z} \cdot \mathbf{\Sigma}) \tau z] \mathbf{a}(0). \]
This is to say that the rotation of the polarization bases in the chiral medium is the reflection of the rotation of the local coordinate system about the propagation direction. 
It is particularly pointed out that since the rotation angle $\tau z$ there is proportional to the propagation distance, the state of polarization at $z=z_1$ is different from the state of polarization at a different propagation distance, say, $z=z_2$, even when $\tau (z_2 -z_1) =2m \pi$ with $m$ a non-zero integer.
More interestingly, a helically-coiled fiber that is free of linear birefringence not only rotates the transverse axes of the local coordinate system but also changes the propagation direction \cite{Papp-H, Ross, Qian-H, Chen-R} if a locally-paraxial approximation is made to the wave that passes through it. 

The difference between the new and the traditional mechanisms can be further explained as follows. 
It is seen from Eq. (\ref{a''}) that the traditional mechanism is mathematically expressed as an SU(2) rotation of the Jones vector within a local coordinate system. The corresponding Poincar\'{e} vector undergoes an SO(3) rotation in the same local coordinate system, as is indicated by Eq. (\ref{RoPV}).  
However, as is shown by Eq. (\ref{a'-a}), the new mechanism leads to an SO(3) rotation of the polarization vector in the laboratory coordinate system. What is important is that the Poincar\'{e} vector in this case cannot always embody such a rotation even though it can be viewed as a vector in the laboratory coordinate system. 
To illustrate this more clearly, let the state of quasi-spin be an eigenstate of $\hat{\sigma}_3$, say, 
$
\alpha=\frac{1}{\sqrt 2} \bigg(\begin{array}{c}
	                              1 \\ i
                               \end{array} 
                         \bigg)
$.
As we have just seen, the polarization state,
$\mathbf{a}' =\frac{1}{\sqrt 2}(\mathbf{u}' +i \mathbf{v}')$,
in association with the primed local coordinate system $\varpi'$ is different from the polarization state,
$\mathbf{a}=\frac{1}{\sqrt 2}(\mathbf{u}+i \mathbf{v})$,
in association with the unprimed local coordinate system $\varpi$.
As a matter of fact, since $\mathbf{u}+i \mathbf{v}$ is an eigen vector of the operator
$\mathbf{w} \cdot \mathbf{\Sigma}$ with the eigenvalue $+1$, it follows from Eq. (\ref{a'-a}) that $\mathbf{a}'$
is different from $\mathbf a$ by a phase factor,
\[ \mathbf{a}' =\exp(-i \phi) \mathbf{a}. \]
But the Poincar\'{e} vectors in both local coordinate systems are the same and take the form of $\mathbf w$. 
This again shows that the Poincar\'{e} vector is not able to completely characterize the polarization of a plane light wave.
The key point here is that the Poincar\'{e} vector, when viewed in the laboratory coordinate system, generally has a non-vanishing longitudinal component, in sharp contrast to the polarization vector that has no longitudinal component.

\section{Conclusions and Remarks}\label{conclusions}

In conclusion, realizing that the traditional description of the polarization of light is incomplete, we reexamined the nature of the polarization from the point of view of measurement.
We found that in order to completely determine the polarization state of a plane light wave, two different kinds of degrees of freedom are needed. One is the Stokes parameters. The other is the degrees of freedom to specify the local coordinate system in which the Stokes parameters are defined.
This is because described by the polarization vector in the laboratory coordinate system, the polarization of a light wave is the reflection of its quantum-mechanical property, the quasi-spin, in the momentum-dependent local coordinate system.
The wavefunction of the quasi-spin is the Jones vector. As is indicated by (\ref{alpha-a}), the Jones vector is the projection of the polarization vector onto the local coordinate system represented by the matrix $\varpi^\dag$. It appears as a spinor in the local coordinate system. 
The representative operator for the quasi-spin is given by (\ref{QSO}). The SU(2) algebra (\ref{su2}) makes the Stokes parameters (\ref{SP}) form a vector, the Poincar\'{e} vector (\ref{PV}), in the local coordinate system. 
The Poincar\'{e} vector, or equivalently, the Stokes parameters, characterizes the state of quasi-spin in accordance with (\ref{EVE}).

On the one hand, the state of quasi-spin in a definite local coordinate system has a one-to-one correspondence with the state of polarization in the laboratory coordinate system. In this case, the Stokes parameters can be used to differentiate between different states of polarization. This is what is done in traditional description.
On the other hand, a definite state of quasi-spin in different local coordinate systems corresponds to different states of polarization. The Stokes parameters in such a case can no longer be used to distinguish different states of polarization.
Consequently, there are two distinct mechanisms to change the state of polarization without affecting the intensity. One is to change the state of quasi-spin, or the Stokes parameters, in a fixed local coordinate system. The other is to change the local coordinate system with the Stokes parameters remaining fixed in it. The former is the traditional mechanism that can be expressed as an SU(2) rotation of the Jones vector. The latter, which is newly identified here, is found to be responsible for the optical activity in the chiral medium.

In a word, the Stokes parameters by themselves are not able to completely characterize the polarization of a pure state in classical optics.
This result provides a deeper insight into the nature of optical polarization. It is closely related to the ``hidden polarization'' \cite{Bjork} that was conjectured a long time ago \cite{Klys,Kara-93}. 
It should be pointed out that the quasi-spin of the photon is not to be confused with the spin angular momentum that combines the orbital angular momentum to give the total angular momentum \cite{Enk-N, Li}.

\section*{Acknowledgments}

This work was supported in part by National Natural Science Foundation of China under Grant No. 11974251.


\begin{thebibliography}{99}
\bibitem{Comm1}
Eq. (\ref{rotation-JV}) is not to be confused with Eq. (\ref{trans-JV}), which expresses the transformation of the Jones vector of one given polarization vector under the transformation of the local coordinate system.
\bibitem{Comm2}
This, in fact, is guaranteed by the one-to-one correspondence between the state of polarization in the laboratory coordinate system and the state of quasi-spin in a fixed local coordinate system.
\end{thebibliography}
\end{document}